\documentclass{ws-ijmpa}
\usepackage[super,compress]{cite}
\usepackage{graphicx}
\begin{document}
\markboth{Zhiyang Yuan, Huirong Qi}{Feasibility study of TPC tracker detector for the circular collider}

%
\catchline{}{}{}{}{}
%

\title{Feasibility study of TPC tracker detector for the circular collider}

\author{Zhiyang Yuan, Huirong Qi\footnote{Huirong Qi, qihr@ihep.ac.cn}, Haiyun Wang, Ling Liu, Yuanbo Chen, Qun Ouyang, Jian Zhang}

\address{State Key Laboratory of Particle Detection and Electronics,\\
Institute of High Energy Physics, Chinese Academy of Sciences,\\
19B Yuquan Road, Beijing 100049, China\\
University of Chinese Academy of Sciences,\\
19A Yuquan Road, Beijing 100049, China\\
}

\author{Yiming Cai, Yulan Li}

\address{Department of Engineering Physics, Tsinghua University,\\
Hai Dian District, Beijing 100084, China\\
qihr@ihep.ac.cn}

\maketitle

\begin{history}
\received{Day Month Year}
\revised{Day Month Year}
\end{history}

\begin{abstract}
The discovery of a SM Higgs boson at the LHC brought about great opportunity to investigate the feasibility of a Circular Electron Positron Collider (CEPC) operating at center-of-mass energy of $\sim 240$ GeV, as a Higgs factory, with designed luminosity of about $2\times 10^{34}cm^{-2}s^{-1}$. The CEPC provides a much cleaner collision environment than the LHC, it is ideally suited for studying the properties of Higgs boson with greater precision. Another advantage of the CEPC over the LHC is that the Higgs boson can be detected through the recoil mass method by only reconstructing Z boson decay without examining the Higgs decays. In Concept Design Report(CDR), the circumference of CEPC is 100km, with two interaction points available for exploring different detector design scenarios and technologies. The baseline design of CEPC detector is an ILD-like concept, with a superconducting solenoid of 3.0 Tesla surrounding the inner silicon detector, TPC tracker detector and the calorimetry system. Time Projection Chambers (TPCs) have been extensively studied and used in many fields, especially in particle physics experiments, including STAR and ALICE. The TPC detector will operate in continuous mode on the circular machine. To fulfill the physics goals of the future circular collider and meet Higgs/$Z$ run, a TPC with excellent performance is required. We have proposed and investigated the ions controlling performance of a novel configuration detector module. The aim of this study is to suppress ion backflow ($IBF$) continually. In this paper, some update results of the feasibility and limitation on TPC detector technology R$\&$D will be given using the hybrid gaseous detector module.
\end{abstract}

\ccode{PACS numbers:29.40.Cs}


\section{Introduction}	
Higgs boson at the LHC\cite{lhc} brought about great opportunity to investigate the feasibility of a Circular Electron Positron Collider (CEPC)\cite{cepc} operating at center-of-mass energy of $\sqrt s=$~240 GeV, as a Higgs factory, with designed luminosity of about $2\times 10^{34}cm^{-2}s^{-1}$. The CEPC provides a much cleaner collision environment than the LHC, it is ideally suited for studying the properties of Higgs boson with greater precision. It also provides the best probe into the Higgs invisible decays and search for dark matter and exotic particles produced in the Higgs decays. The CEPC can also operate at the $Z$ pole and near the $WW$ threshold to allow for refined measurement of the SM parameters with significantly higher precision. The circumference of CEPC will be 100km, with two interaction points available for exploring different detector design scenarios and technologies. Two common RF stations are deployed for the Higgs operation, which result in 286 beam bunches evenly distributed over a half ring. While for $W$ and $Z$ operations, independent RF cavities are used, 5220 and 10900 bunches are spreading in equal distance over the full ring, respectively. Therefore, the bunch spacing are about 500ns, 50ns and 30ns for Higgs, $W$ and $Z$ operations, respectively.

The baseline design of CEPC detector is an ILD-like concept\cite{ild}, with a superconducting solenoid of 3.0 Tesla surrounding the inner silicon detector, TPC tracker detector and the calorimetry system. In order to accommodate the CEPC collision environment, some necessary changes have been made to the Machine Detector Interface (MDI) and sub-detector design. The CEPC design, for instance, has a significantly shorter focal length $L\*$ of 2.2m than that of the ILC design (3.5m), which indicates that the final focusing magnet $QD_0$ will be placed inside the CEPC detector. In addition, unlike the ILC detector, the CEPC detector will operate in continuous mode\cite{ibf}, which imposes special considerations on power consumption and subsequent cooling of the sub-detectors.

Time Projection Chambers (TPCs)\cite{tpc} have been extensively studied and used in many fields, especially in particle physics experiments, including STAR\cite{star} and ALICE\cite{alice}. Their low material budget and excellent pattern recognition capability make them ideal for three dimensional tracking and identification of charged particles. The TPC detector will operate in continuous mode on the circular machine. To fulfill the physics goals of the future circular collider and meet Higgs$/Z$ run, a TPC with excellent performance is required. MPGDs with outstanding single-point accuracy and excellent multi-track resolution are needed. We have proposed and investigated the ions controlling performance of a novel configuration detector module. The aim of this study is to suppress ion backflow ($IBF$) continually.

\section{Continuous $IBF$ suppression TPC module of $GEM\!\!-\!\!MM$}
There has been a critical problem with TPC detector, especially in high background conditions – the space charge distortion due to the accumulation of positive ions in the drift volume. Due to their large mass, positive ions move slowly under the action of electric field in the drift volume of the TPC. The continuously superimposed ions in the drift volume of the TPC may affect the drift behaviour of electrons from a later track. The majority of ions inside the drift volume are back flowing ions from the amplification region of the TPC readout devices. It is thus of great importance to limit ion backflow from the amplification region. Early TPCs were equipped with multi-wire proportional chambers (MWPCs)\cite{mwpc} as gas amplification devices. The $IBF$ ratio in a standard MWPC is $30\%-40\%$, so a gating GEM is essential to prevent ions from reaching the drift volume.

Under $Z$ pole run mode in the circular machine, there is the not enough the open/close time for this technology option. The idea of combining GEM with Micromegas was first proposed with the goal of reducing the spark rate of Micromegas detectors. Pre-amplification using GEM also extends the maximum achievable gain, so there have also been studies on gaseous photomultipliers with this hybrid configuration. The cascaded structure of the $GEM\!\!-\!\!MM$ detector(GEM with Micromegas) is composed of a drift electrode, a GEM foil, a standard Micromegas, and a readout printed circuit board. The Micromegas detector is based on the bulk method and has an active area of $100mm\times100mm$. The micromesh is made of stainless steel wires 22$\mu m$ in diameter, interwoven at a pitch of 62$\mu m$. 128$\mu m$ under the micromesh is a single copper pad readout plane. A GEM foil is cascaded above the micromesh at a distance of 1.4$mm$. It is a standard GEM foil of area $100mm\times100mm$, produced from CERN. In the experiment, the drift distance was maintained at 4$mm$. Electrodes were biased with CAEN N471A\cite{caen} high voltage units.

$^{55}Fe$ source was used to produce the primary electrons in the sensitive volume during the test. The working gas was a mixture of $Ar/CO_2\!\!=\!\!90/10$ and $Ar/CF_4/iC_4H_{10}\!\!=\!\!95/3/2 (T2K)$ operating gases at the room temperature and the atmospheric pressure. Ion backflow is due to secondary ions generated in an electron-avalanche process in the amplification which return to the drift space. In this paper, fractional ion feedback is defined as the ratio of the ion charge injected into the drift volume, collected on the drift electrode, and the electron charge collected on the anode pad. A Keithley (6517B)\cite{6517b} pico current electrometer device was used to measure the current with a resolution of about $1.0pA$.

\begin{figure}[!htb]
\centerline{\includegraphics[width=11.5cm]{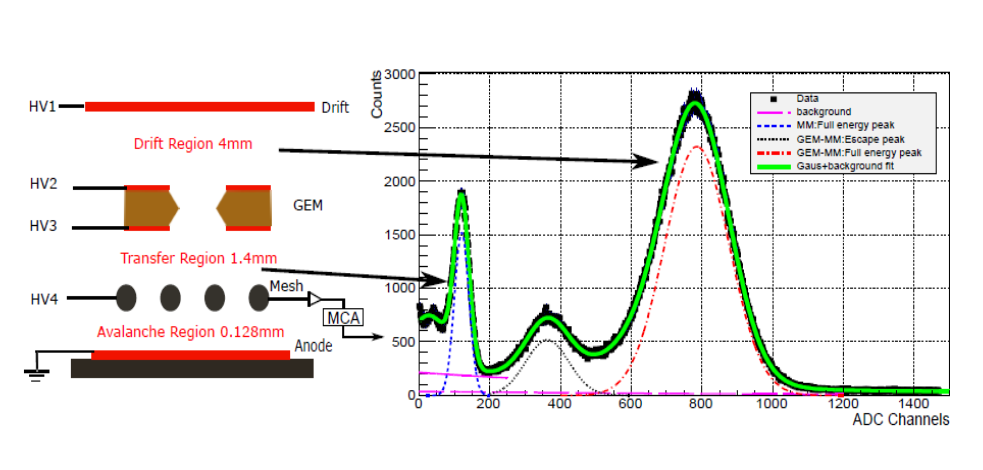}}
\caption{$GEM\!\!-\!\!MM$ configuration (left) and energy spectrum at 5.9 keV@$^{55}Fe$, showing each peak and the corresponding location of primary ionization (right). Source: $^{55}Fe$; Gas: $Ar/CO_2\!\!=\!\!90/10$;$E_d$=250V/cm, $V_{GEM}$=340V, $E_t$=500V/cm, $V_{mesh}$=420V. \label{f1}}
\end{figure}
\subsection{Energy spectrum with $GEM\!\!-\!\!MM$ detector}

$^{55}Fe$  X-ray source with a characteristic energy of 5.9keV was used in the test. In the argon-based working gas mixture, a typically pulse height spectrum for a GEM or Micromegas detector contains one major peak corresponding to the 5.9keV X-rays and an escape peak at lower pulse heights corresponds to the ionization energy of an electron from the argon $K$-shell. In the $GEM\!\!-\!\!MM$ detector, the situation is different. There are two amplification stages inside this detector. The primary ionization created by photon absorption can be in the drift region or in the transfer region. Photoelectrons starting from the drift region get amplified by both the GEM detector and the Micromegas detector before they are collected on the anode. If the photons are absorbed in the transfer region, the primary electrons will be amplified only once (by Micromegas).

Figure.1(see Fig.~\ref{f1}) depicts a typical $^{55}Fe$  pulse height spectrum obtained by the $GEM\!\!-\!\!MM$ detector. Four peaks are seen in the pulse height spectrum. From left, the first peak and the second peak are the escape peak and the full energy peak of the stand alone Micromegas. The last two peaks are created by photons with their energy deposited in the drift region. These primary electrons show combination amplification. The principle of the $GEM\!\!-\!\!MM$ detector is fully verified.

\begin{figure}[!htb]
\centerline{\includegraphics[width=8.5cm]{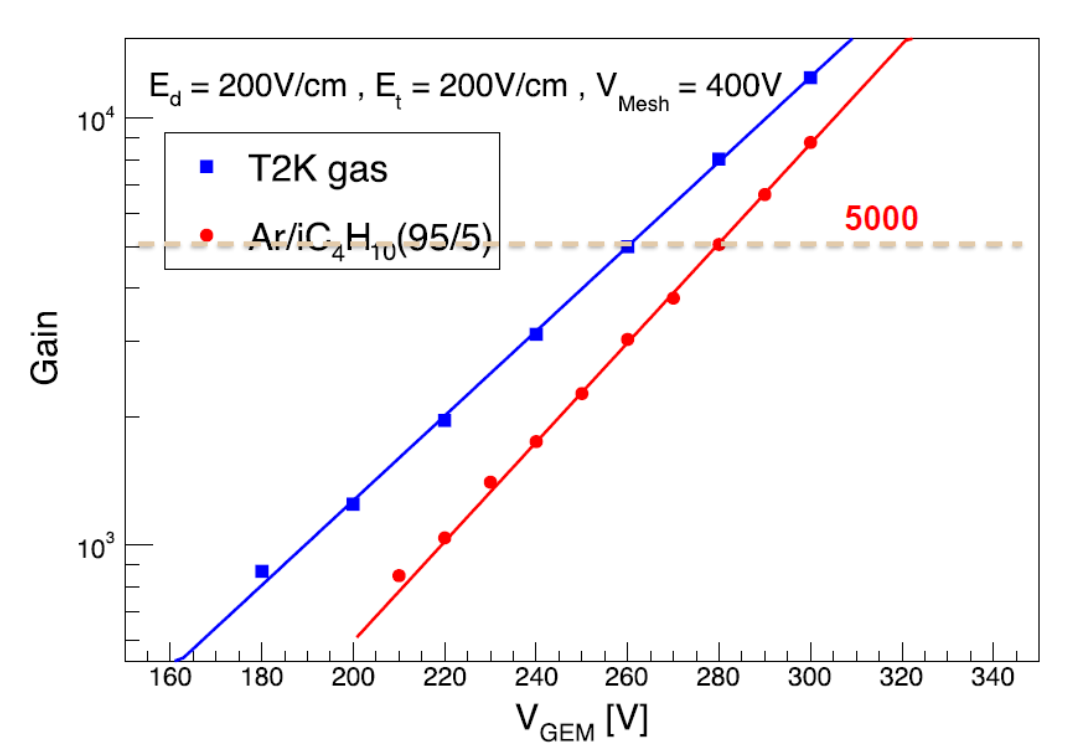}}
\caption{Gas gain at the mixture operation gases of $T2K$ and $Ar/iC4H_{10}\!\!=\!\!95/5$. \label{f2}}
\end{figure}
\subsection{Relative Gain}

With the calibrated electronic gain results, the gain of the detector is obtained from the measured spectra as described in the previous subsection. The gain of the Micromegas or GEM is characterised by the first full energy peak in the spectrum. The last full energy peak represents the overall gain of the $GEM\!\!-\!\!MM$ detector. It is important to note that this is a new way to measure the effective gas gain of a GEM. A gain of 5000 or more can be achieved without any obvious discharge behaviour.

The update gain results shows in Figure.2(see Fig.~\ref{f2}) and the gain of $GEM\!\!-\!\!MM$ detector operated in the two different mixture gas of $T2K$ and $Ar/iC_4H_{10}\!\!=\!\!95/5$ respectively. It indicated that the yellow dash line is the level of 5000 when the gain could reach.

\subsection{Ion Backflow rate}
$IBF$ increases initially and decreases afterwards as the GEM voltage increases. So, an ion backflow value of about $3\%$ is considered to be the $IBF$ for a standalone Micromegas detector with a gain of about $600$. When the GEM is cascaded, the $IBF$ can be further reduced to below $1\%$. Figure.2 shows that when a constant bias voltage is set across the GEM, the $IBF$ decreases as the micromesh voltage increases. The reason is that electrons collected on the anode increase with the increase of the mesh voltage. So the $IBF$ can be estimated as a few percent for a single GEM detector with a comparatively low gain of approximately 4. After the Micromegas is cascaded, $IBF$ is reduced significantly.

\begin{figure}[!htb]
\centerline{\includegraphics[width=8.5cm]{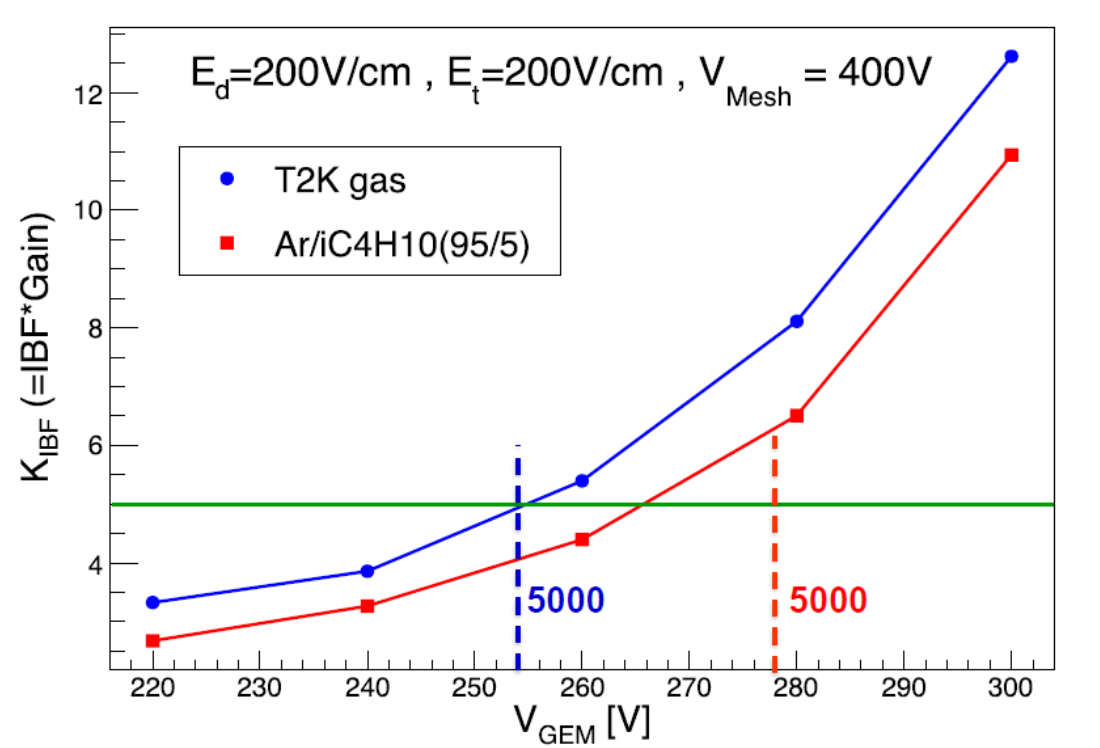}}
\caption{Relative factor of the detector gain and $IBF$ rate. \label{f3}}
\end{figure}

In 2018, the parameters of the electric field of drift, transfer, GEM detector and Micromegas detector have been optimized testing in Institute of High Energy Physics. The key factor of the gas gain times $IBF$ obtained at the mixtures gases of $T2K$ and $Ar/iC_4H_{10}\!\!=\!\!95/5$ separately. The new results has been shown in Figure.3(see Fig.~\ref{f3}) and some test parameters were given. To meet the $Z$ pole run mode in the circular machine, some preliminary results of the simulation and estimation have been done in the publication paper. The TPC detector at the proposed circular collider will have to be operated continuously and the backflow of ions must be minimized without the open/close time of a gating device technology. The currents on the anode and drift cathode were measured precisely with an electrometer, and the experimental results showed that $IBF$ rate can be reduced to about $0.1\%$ at the gain of about 5000.

\begin{figure}[!htb]
\centerline{\includegraphics[width=12.5cm]{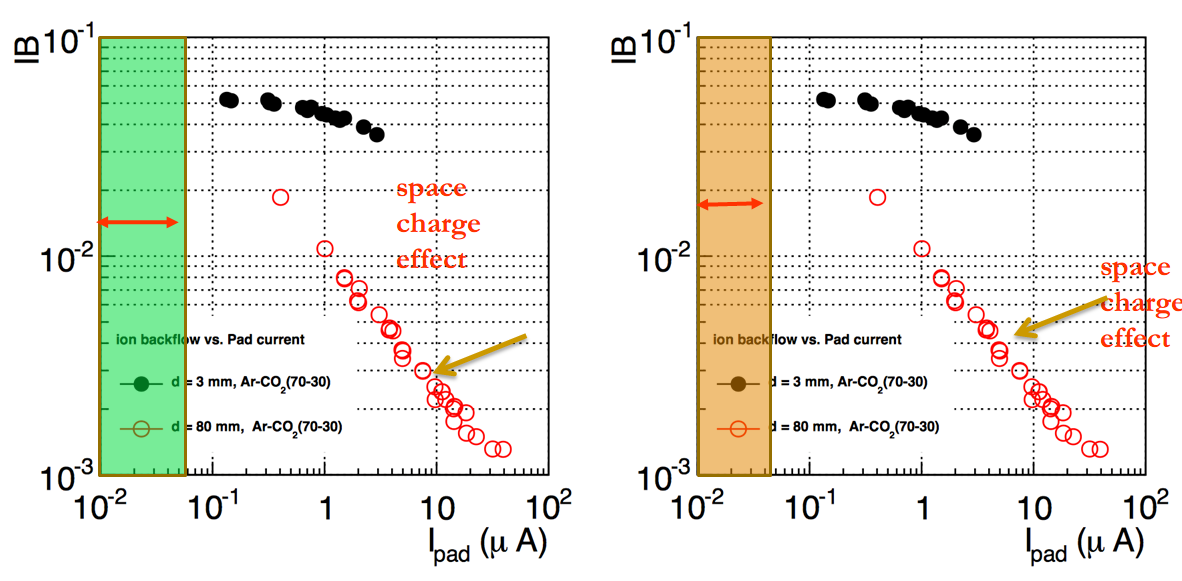}}
\caption{Comparison of the $IBF$ with the different X-ray’s voltage and current. The results of the $GEM\!\!-\!\!MM$ detector module appear in the green/yellow areas without space charge effect. \label{f4}}
\end{figure}

Comparing with the ALICE TPC upgrate detector option with the cascade of four GEMs\cite{alicegems}, the hybrid detector could obtain clear the full photo peak and the electron escape peak, and it shows that the good electron transmission and good energy resolution of less than $20\%$ at 5.9KeV of X-ray. Meanwhile, a lots of ions and electrons make obviously the space charge effect to decrease the $IBF$ rate\cite{spacecharge}. Concerning the update results, the parameter of gain and current of the readout pad have been checked with the different X-ray’s voltage and current, it indicated that there is no space charge effect to decrease the $IBF$ possibility(see Fig.~\ref{f4}) under the two different operation mixture gases of $T2K$(Green area) and $Ar/iC_4H_{10}\!\!=\!\!95/5$(Yellow area) separately. The $IBF$ rate is the initial ions suppression function of the hybrid structure gaseous detector without the space charge effect in our experiment. The preliminary IBF results show that the feasibility module concept to meet on $Z$ pole run in the circular collider, and the space resolution will be studied with the electron beams.

\section{Conclusion}
The Time Projection Chamber presented here provides an excellent starting point for the research and development in the context of the CEPC beam environment. Several studies have already been performed and many more are foreseen. Possible solutions to these issues have been suggested and will continue to be investigated with the TPC prototype. A hybrid-structure MPGD detector module has been developed, and preliminary results have been obtained and analyzed. Further studies will be done from this combination detector module.



\section*{Acknowledgments}

The author thanks for Prof. Yuanning Gao, Prof. Yulan Li and Dr. Yiming Cai for some details discussions. This study was supported by National Key Programme for $S\&T$ Research and Development (Grant NO.: 2016YFA0400400), the National Natural Science Foundation of China (Grant NO.: 11675197) and the National Natural Science Foundation of China (Grant NO.: 11775242)


\end{document}